\documentclass[%
 aip,
 amsmath,amssymb,
reprint,%
]{revtex4-1}

\usepackage{graphicx}
\usepackage{dcolumn}
\usepackage{bm}

\usepackage[utf8]{inputenc}
\usepackage[T1]{fontenc}
\usepackage{mathptmx}
\usepackage{epstopdf}
\usepackage{soul}
\usepackage[colorlinks=true,linkcolor=blue,urlcolor=blue,citecolor=blue]{hyperref}

\begin{document}

\preprint{AIP/123-QED}

\title{Resolution-enhanced quantitative spectroscopy of atomic vapor in optical nanocells based on second-derivative processing of spectra}

\author{A. Sargsyan}
 \affiliation{ Institute for Physical Research, NAS of Armenia, Ashtarak, 0203, Armenia}
\author{A. Amiryan}
 \email{arevmiryan@gmail.com.}
\affiliation{ Institute for Physical Research, NAS of Armenia, Ashtarak, 0203, Armenia}
\affiliation{Laboratoire Interdisciplinaire Carnot de Bourgogne, Universit\'e Bourgogne Franche-Comt\'e,  21078 Dijon Cedex, \looseness=-1{France}}
\author{Y. Pashayan-Leroy}
\affiliation{Universit\'e Bourgogne Franche - Comt\'e, 32 avenue de l’Observatoire, 25000 Besan\c{c}on, \looseness=-1{France}}
\author{C. Leroy}
\affiliation{Laboratoire Interdisciplinaire Carnot de Bourgogne, Universit\'e Bourgogne Franche-Comt\'e, 21078 Dijon Cedex, \looseness=-1{France}}
\author{A. Papoyan}
\affiliation{ Institute for Physical Research, NAS of Armenia, Ashtarak, 0203, Armenia}
\author{D. Sarkisyan}
\affiliation{ Institute for Physical Research, NAS of Armenia, Ashtarak, 0203, Armenia}

\date{\today}

\begin{abstract}
We present a method for recovery of narrow homogeneous spectral features out of broad inhomogeneous overlapped profile based on second-derivative processing of the absorption spectra of alkali metal atomic vapor nanocells. The method is shown to preserve the frequency positions and amplitudes of spectral transitions, thus being applicable for quantitative spectroscopy. The proposed technique was successfully applied and tested for: measurements of hyperfine splitting and atomic transition probabilities; development of an atomic frequency reference; determination of isotopic abundance; study of atom--surface interaction; and determination of magnetic field-induced modification of atomic transitions frequency and probability. The obtained experimental results are fully consistent with theoretical modeling. 
\end{abstract}

\maketitle

 Optical cells containing atomic vapor of alkali metals are widely used in atomic physics, laser spectroscopy, and emerging applications, including chip-scale atomic clocks and optical magnetometry \cite{Kitching}, slowing the light speed and formation of narrow optical resonances \cite{Khurgin}, ultra-narrow optical filters \cite{Tao}, etc. For many processes, such as resonant atom-light interaction in the presence of magnetic or electric fields and collisional broadening of spectral lines, frequency resolution of individual optical transitions between the hyperfine sublevels, which can be separated from each other by down to 20 - 50 MHz, is important. Meanwhile inhomogeneous Doppler broadening in alkali vapor cells reaches 400 - 1500 MHz \cite{Demtroder}. Sub-Doppler resolution with a simple single-beam geometry, providing a linear response of atomic media for transmission, fluorescence, and selective reflection experiments can be attained using optical vapor nanocells with a thickness of the order of resonant wavelength \cite{sarkisyan2004spectroscopy}. But even in this case some spectral features can be smeared by a residual Doppler-overlapped profile \cite{Das}.
 
 We propose a simple data processing technique, which allows retrieval of homogeneous lineshape of individual optical transitions in absorption spectra, which are initially masked by overlapped inhomogeneous spectral profile. The technique is efficient for the spectra of an atomic vapor nanocell (NC) with a vapor column thickness equal to half of the wavelength of the resonant laser radiation \cite{Dutier, Pashayan, Amiryan}. The following procedure is employed. After recording the absorption spectrum with scanning the laser radiation frequency across the resonant region, the second order derivative  $A''(\omega)$ is determined from the raw absorption  spectrum $A(\omega)$. As a result, the transition linewidth in the second derivative (SD) spectrum reduces down to $\sim$ 15 MHz ($\sim$ 50-fold narrower as compared with the Doppler broadening). This allows to separate and study individual atomic transitions. Here we demonstrate that the SD treatment of the recorded signal preserves not only frequency positions of spectral features but, importantly, also their amplitude values linked, in linear interaction regime, with transition probabilities. We present below some particular realizations of resonant optical processes where the SD technique is employed, and prove its capability for quantitative measurements. 
 
 \begin{figure}
 	\includegraphics[width=7.0cm,keepaspectratio=true]{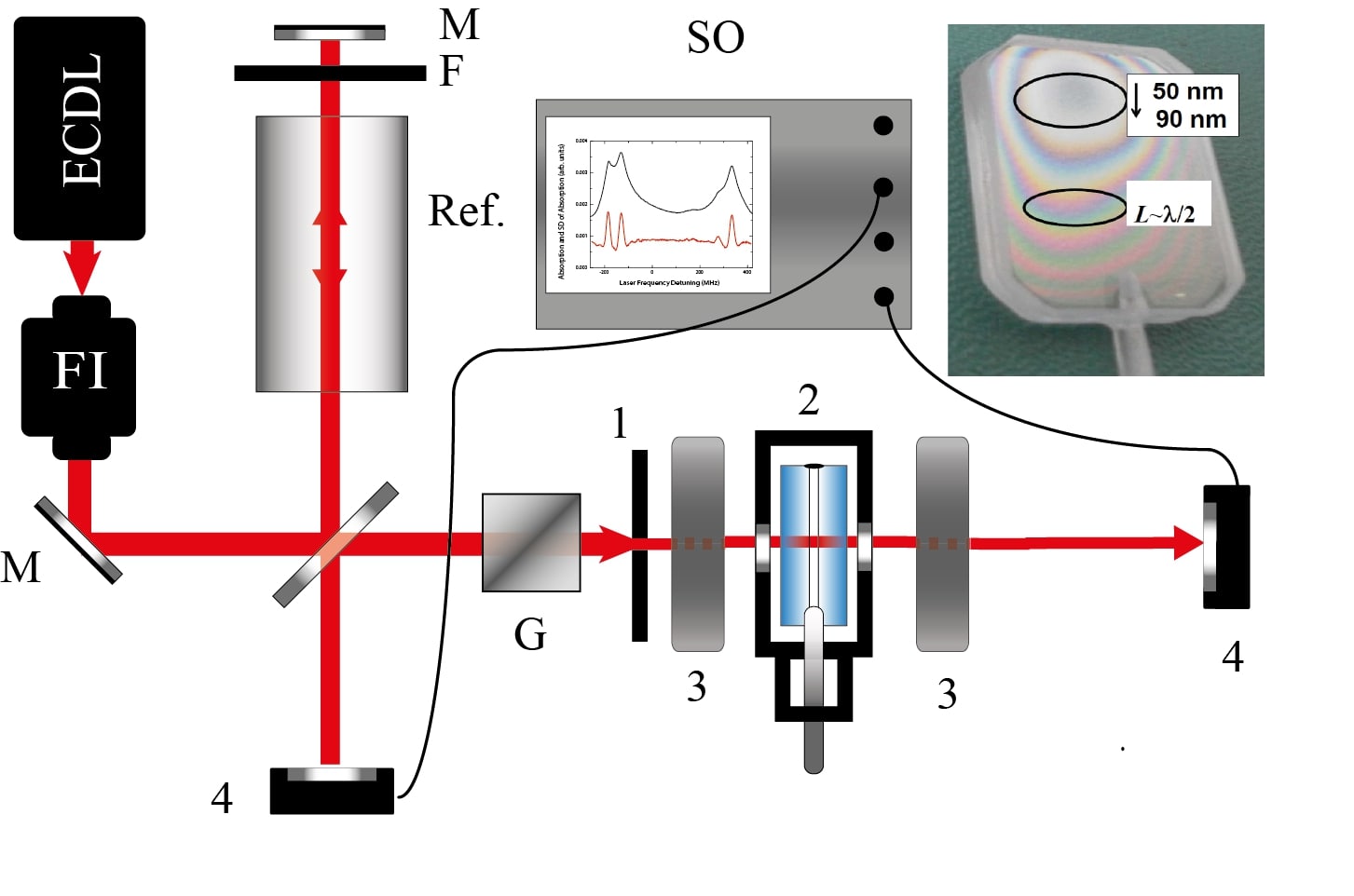}
 	\caption{\label{fig:Fig1} Sketch of the experimental setup. ECDL-diode laser; FI - Faraday isolator; 1 - $\lambda/4$ plate (optional); 2 - NC filled with Rb or Cs or K in the oven; 3 - permanent magnets; 4 - photodetectors; M - mirror; F - filter; SO - digital storage oscilloscope; G - Glan polarizer; Ref. - frequency reference unit. Inset: photograph of the NC filled with Cs; the upper and lower ovals mark the thickness regions 50 - 90 nm and $\sim$ 426 nm in the case of Cs ($L \approx \lambda/2$), respectively.}
 \end{figure}
 
The experimental setup is schematically sketched in Fig.\ref{fig:Fig1}. Tunable external-cavity diode lasers (ECDL) with spectral linewidth $<$ 1 MHz and wavelength $\lambda$ = 780, 852, 767, and 770 nm were used for resonant excitation of atomic vapor on $D_2$ lines of Rb, Cs, K and $D_1$ line of K, respectively. The laser beam was directed normally to the windows of NC containing atomic vapor with a thickness of atomic vapor column $L \sim \lambda/2$. The details of NC design are presented elsewhere \cite{Keaveney}. The absorption spectra are recorded with the help of photo-diodes, and a four-channel digital storage oscilloscope Tektronix TDS2014B. Thanks to the use of NC, a uniform magnetic field variable in the range of $B$ = 10 -- 10000 G can be applied to the vapor by strong permanent magnets mounted on the translation stages \cite{Sargsyan}.
 
\begin{figure}
\includegraphics[width=8.0cm,keepaspectratio=true]{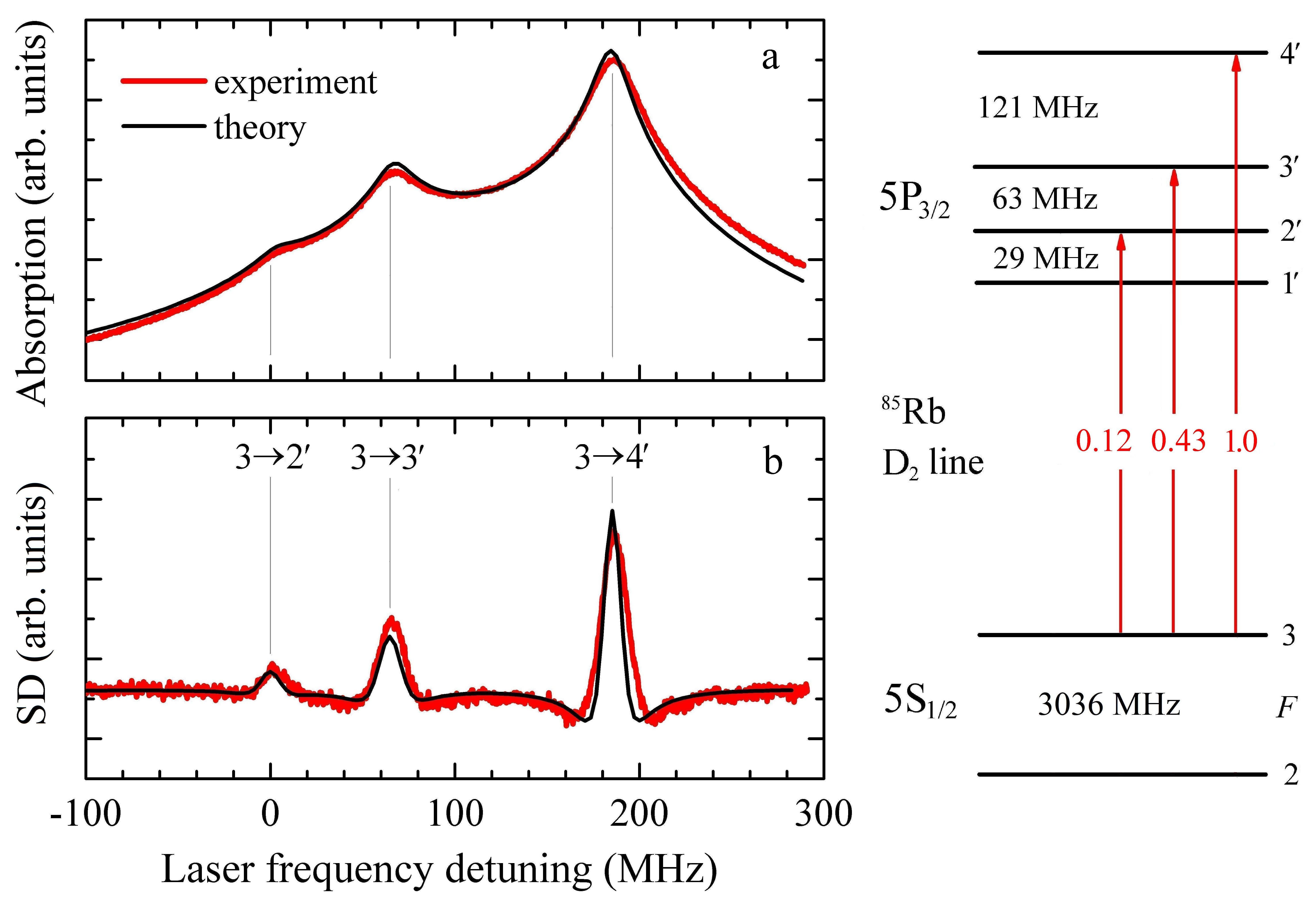}
\caption{\label{fig:Fig2} a) Absorption spectrum of a rubidium NC with $L \sim 390$ nm for $3 \rightarrow$ 2$'$,3$'$,4$'$ hyperfine transitions of $^{85}$Rb $D_2$ line. b) Second derivative of the absorption spectrum. Red lines: experiment; black lines: theory. Gray vertical lines indicate frequency positions of the transitions. Right panel: energy levels diagram, showing the transitions, their frequency separations and relative probabilities. Here and bellow the probabilities are normalized to the probability of the strongest transition in the group.}
\end{figure}

The first verification of the proposed technique was done for the NC absorption spectra on Rb $D_2$ line, for hyperfine transitions $F_g=3 \rightarrow F_e=2,3,4$ (hereafter we will use a simplified notation 3 $\rightarrow$ 2$'$,3$'$,4$'$, see the right panel in Fig.\ref{fig:Fig2}). These transitions are poorly resolved even in the NC absorption spectra, as is seen from Fig.\ref{fig:Fig2}a, meanwhile they are completely resolved in the SD spectrum (Fig.\ref{fig:Fig2}b) due to the small ($\sim$ 15~MHz FWHM) linewidth. Here and below, the modeling of experimental spectra was done using the previously developed theoretical model \cite{Dutier, Pashayan, Amiryan} with the following key assumptions: the vapor density $N$ is low enough, so that the effect of interatomic collisions can be is ignored, and only atom–surface collisions are considered; the atoms experience inelastic collisions with the NC walls. The effect of reflection of the laser radiation from both highly-parallel windows of the NC, as well as Doppler-broadening effect are taken into account. In the experiment, the temperature of the NC reservoir was kept at 110~$^\circ$C corresponding to vapor density $N_{Rb} \approx$ 1.1$\times$10$^{13}$ cm$^{-3}$, the laser radiation power was $P_L$ = 10 $\mu$W, sufficiently low to avoid saturation and optical pumping effects. The extra broadening over the natural linewidth $\gamma_N/2\pi = 6$ MHz is caused by the atom-surface collisions.\\
\indent As one can see from Fig.\ref{fig:Fig2}b, the SD method allows retrieving the homogeneous absorption spectra from inhomogeneously (e.g. Doppler) broadened one, with the use of a single low-power laser beam. Moreover, the SD method correctly displays both the frequency positions of individual transitions and their relative probabilities (the inaccuracy is 3-5\%). It is also important to note that precise value $L = \lambda/2$ for the NC thickness is not crucial: narrow spectra are observed with $\Delta L = \pm$ 50 nm tolerance, which makes utilization of the proposed technique experimentally feasible.\\
\indent Relatively narrow unshifted resonances obtained with SD method allow implementation of the SD spectra for frequency reference (FR) applications. A FR technique mostly used in atomic physics experiments is based on saturated absorption (SA) spectroscopy \cite{Demtroder}. In this technique, the laser beam is split into a weak probe field and a strong pump field, which are sent to the interaction cell as overlapping counter-propagating beams. This allows to form Doppler-free velocity-selective optical pumping (VSOP) resonances located at the line center. When multiple hyperfine transitions are Doppler-overlapped, additional crossover (CO) resonances appear in the spectrum, with the amplitudes, which can exceed the VSOP amplitudes because of strongly nonlinear interaction regime. This may cause distortions affecting "real" atomic resonances.
 
\begin{figure}[h]
\includegraphics[width=7.5cm,keepaspectratio=true]{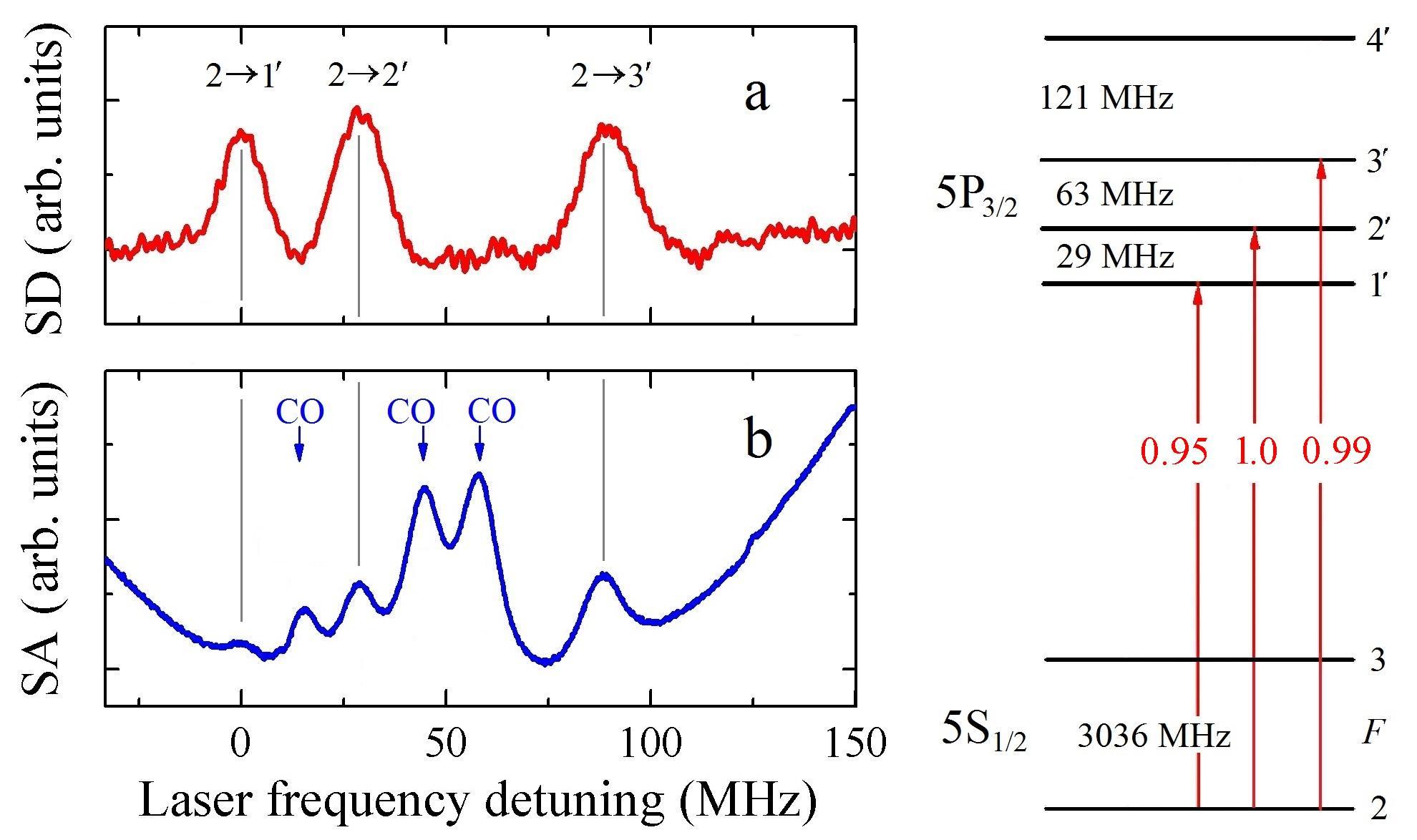}
\caption{\label{fig:Fig3} Experimentally recorded frequency reference spectra formed for 2 $\rightarrow$ 2$'$,3$'$,4$'$ transitions of $^{85}$Rb $D_2$ line using absorption SD spectrum in a NC (a) and SA in an ordinary cell (b). Frequency positions of hyperfine transitions are marked with vertical gray lines; CO labels indicate positions of crossover resonances in SA spectrum. Right panel: energy levels diagram, showing the transitions, their frequency separations and relative probabilities.}
\end{figure}

Comparison of SD and SA spectra recorded for 2 $\rightarrow$ 2$'$,3$'$,4$'$ transitions of $^{85}$Rb $D_2$ line is presented in  Fig.\ref{fig:Fig3}. Only three resolved atomic resonances are observed in the SD spectrum (Fig.\ref{fig:Fig3}a), while for the FR formed using SA technique (Fig.\ref{fig:Fig3}b) there are three VSOP and three CO resonances. The following advantages of the SD method as a FR tool can be mentioned: i) simplicity of the realization geometry (single-beam transmission as opposed to counter-propagating beams requirement for SA geometry); ii) low value of the required laser power (3 orders less than for the SA method); iii) absence of crossover resonances; iv) correspondence of the amplitudes of atomic resonances to their transition probabilities: in particular, the measured ratio of $2 \rightarrow 2'$ and $2 \rightarrow 1'$ SD amplitudes is 1.05, close to the ratio of their probabilities (1.053), while this ratio for the SA method is around 5.

\begin{figure}[h]
\includegraphics[width=7.5cm,keepaspectratio=true]{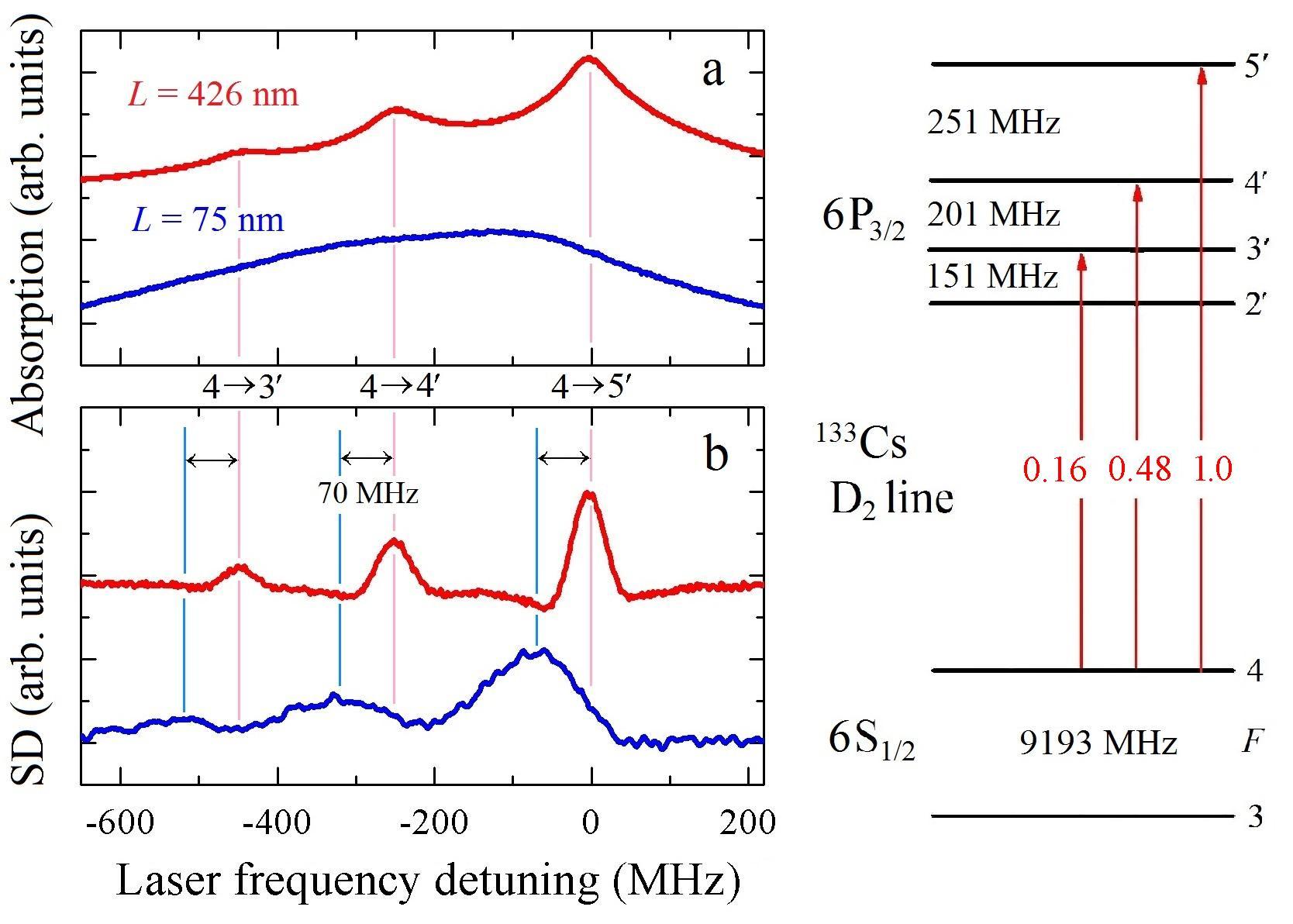}
\caption{\label{fig:Fig4} Experimental absorption (a) and SD (b) spectra for 4 $\rightarrow$ 3$'$,4$'$,5$'$ transitions of $^{133}$Cs $D_2$ line, recorded for two values of the NC thickness marked by ovals on the Cs photograph in Fig.\ref{fig:Fig1}: $L$ = 426 nm (red lines) and $L$ = 75 nm (blue lines). $P_L$ = 10 $\mu$W. A red shift of $\approx$ 70 MHz caused by the vdW interaction is seen for all the frequency-resolved transition components in SD spectra (see text). Right panel: energy levels diagram, showing the transitions, their frequency separations and relative probabilities.}
\end{figure}

Another application where SD method can be employed is the study atom--surface van der Waals (vdW) interaction, which can be revealed from atomic spectra for $L < 100$ nm \cite{Chevrollier, Sargsyan4}, where the use of nanocells is advantageous. Absorption spectra for  4 $\rightarrow$ 3$'$,4$'$,5$'$ transitions of $^{133}$Cs $D_2$ line, recorded for two values of the NC thickness, $L \approx \lambda/2 \approx$ 426 nm and $L$ = 75 nm, are presented in Fig.\ref{fig:Fig4}, together with the corresponding SD spectra. For $L$ = 426 nm, due to the relatively large cell thickness (i.e. large mean distance of atoms from the sapphire windows of NC), the atomic transitions do not undergo frequency shift, thus serving as FR markers \cite{Sargsyan4}. The red shift of frequency is observable for $L = 75$ nm, but the strong broadening and overlapping of individual lines in the absorption spectrum (Fig.\ref{fig:Fig4}a) make quantitative measurements of the frequency shift impossible. Application of SD processing results in complete frequency resolution of hyperfine transitions, revealing $\approx$ 70 MHz vdW red shift for all the three hyperfine transition components due to the interaction. The latter allows to estimate the value of $C_3$ vdW interaction coefficient using expression $\Delta \nu_{vdW} = -16C^3 / L^3$ presented in Ref. \cite{Sargsyan4}, which yields  $C_3$ = 1.8$\pm$0.3  kHz$\times$ $ \mu$m$^3$, which is in a good agreement with the value reported in Ref. \cite{Chevrollier}. The inaccuracy arises from the inaccuracy in determining the NC thickness ($\pm$ 5 nm). We should note that similar measurements can be done for other transitions of Cs, and equally for other alkali metals.

\begin{figure}[h]
	\includegraphics[width=8.0cm,keepaspectratio=true]{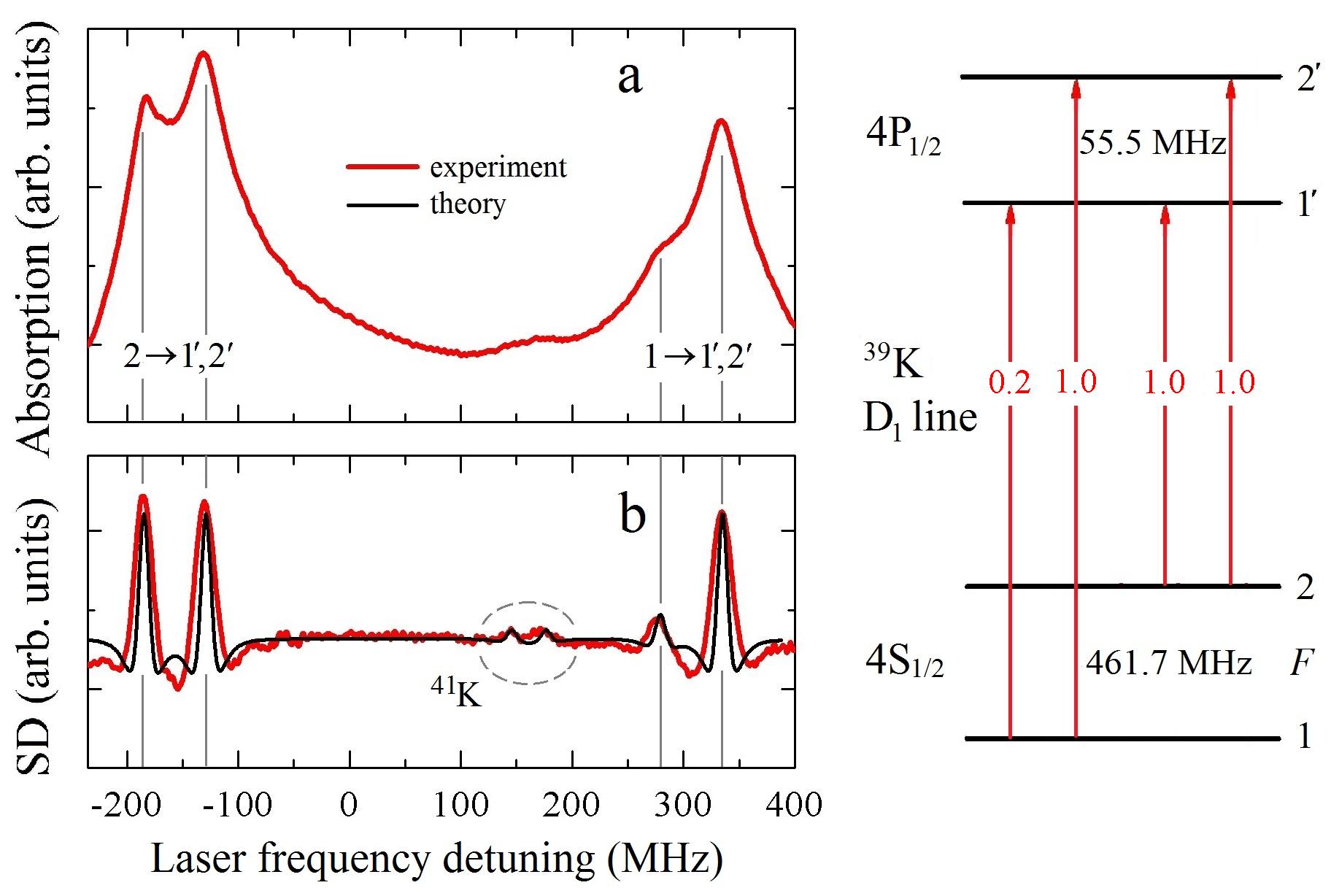}
	\caption{\label{fig:Fig5} Absorption (a) and SD (b) spectra for $D_1$ line of K, recorded in a NC with $L\approx \lambda/2 \approx$ 385 nm, and $P_L$ = 10 $\mu$W. Red lines: experiment; black line: theory. SD peaks corresponding to 2 $\rightarrow$ 1$'$,2$'$ transitions of $^{41}$K separated by 31 MHz are clearly seen (marked by a dotted oval). Right panel: energy levels diagram of $^{39}$K, showing the transitions, their frequency separations and relative probabilities.}
\end{figure}

Furthermore, the SD method can be used for determination of isotopic abundance. To verify this possibility experimentally, we have chosen a NC filled with potassium as an alkali metal with strongly different percentages in natural isotopic mixture, with predominant abundance of $^{39}$K (93.25 $\%$), and minor abundance of $^{41}$K (6.7$\%$). Experiment was done for $D_1$ line of K, using a NC with $L~\approx~\lambda/2~\approx$ 385 nm and reservoir temperature 150~$^\circ$C corresponding to $N_K \approx$ 1.3$\times$10$^{13}$ cm$^{-3}$. The measured absorption and processed SD spectra are shown in Fig.\ref{fig:Fig5}. Thanks to the small linewidth obtained in SD spectrum ($\approx$ 15 MHz, 60 times narrower than the Doppler width), all the hyperfine transitions of $^{39}$K are completely resolved. Besides, the SD spectrum reveals also the hyperfine transitions 2 $\rightarrow $ 1$'$,2$'$ of $^{41}$K $D_1$ line with the frequency separation of 31 MHz (dash-circled in Fig.\ref{fig:Fig5}b). In the weak and linear absorption limit, the amplitude of absorption peak $A$ is proportional to the absorption cross-section $\sigma$, the atomic vapor density $N$, and the atomic vapor thickness $L$: $A~\sim \sigma NL$, see Ref. \cite{Demtroder}. Therefore, the ratio of the amplitudes $A_{^{39}K}/A_{^{41}K} \approx$ 14 for the corresponding hyperfine transitions of $^{39}$K and $^{41}$K isotopes measured from Fig.\ref{fig:Fig5}b must be proportional to the ratio of their partial densities $N_{^{39}K}/N_{^{41}K} \approx$ 13.9 expected from the isotopic abundance. A good agreement between the amplitude values obtained for experimental and modelled spectra (see Fig.\ref{fig:Fig5}b) confirms the validity of the above suppositions concerning the employed absorption regime. 

We should mention that the absorption SD technique can be extended for determination of isotopic composition of other, non-alkali vapor. In Ref. \cite{Witkowski}, SA method was employed to measure the isotopic abundance of mercury (a mixture of five Hg isotopes with the minimum  frequency spacing of 100 MHz) in an atomic vapor cell. Implementation of the SD method for such measurements could be more straightforward. For this purpose it is expedient to use recently developed glass nanocells \cite{Peyrot}, which are simpler in fabrication, but, unlike sapphire NCs \cite{Keaveney}, do not safeguard immunity against highly-aggressive hot alkali vapors.

Correct mapping of resonance frequency and transition probability provided by the SD method makes it attractive when studying processes linked with splitting of atomic transitions and modification of their probabilities in an external magnetic field. The quantity $B_0 = A_{hfs}/\mu_B$ is introduced for quantitative characterization of the atom -- $B$-field interaction \cite{Olsen}, where $A_{hfs}$ is the magnetic dipole constant of the atom’s hyperfine structure, and $\mu _B$ is the Bohr magneton. In strong magnetic fields ($B \gg B_0$), the total electron momentum \textbf{J} and nucleus momentum \textbf{I} become decoupled, manifesting establishment of a hyperfine Paschen–Back (HPB) regime \cite{Olsen, Sargsyan3}, where atomic system is described by momentum projections $m_J$ and $m_I$. 

\begin{figure}[h]
\includegraphics[width=8.8cm,keepaspectratio=true]{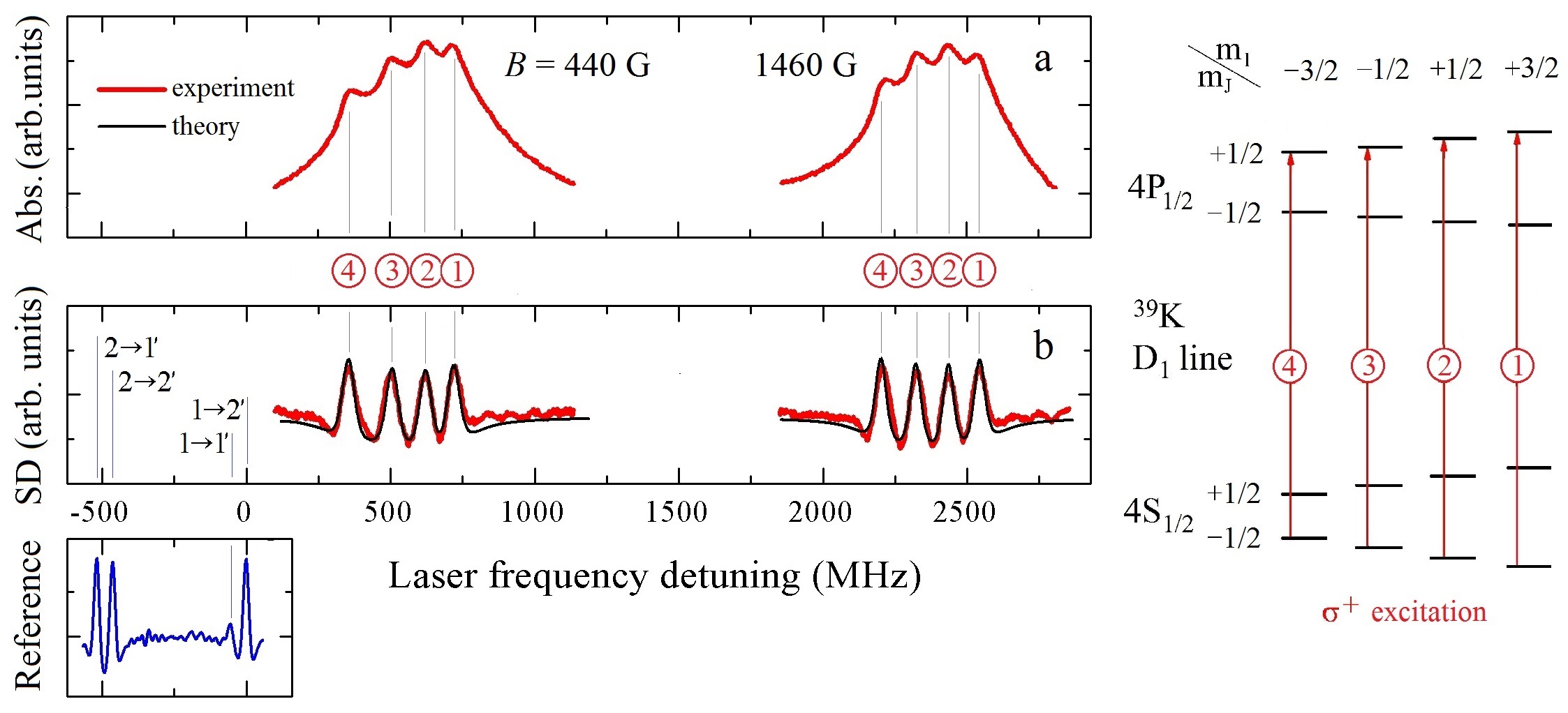}
\caption{\label{fig:Fig6} Absorption (a) and SD (b) spectra for $D_1$ line of $^{39}$K, recorded using $\sigma^+$-polarized excitation in a NC with $L\approx$ 385 nm for two values of a longitudinal magnetic field: $B$ = 440 and 1460 G. Red lines: experiment, black lines: theory. Right panel shows the transitions diagram; the four components remaining in the HPB regime are labeled in circles. The lower left graph (blue line) is the reference spectrum recorded at $B=0$.}
\end{figure}

Absorption and SD spectra exhibiting splitting of $D_1$ line atomic transitions of $^{39}$K in the magnetic field are presented in Fig.\ref{fig:Fig6}. The spectra are recorded with circularly ($\sigma^+$) polarized laser radiation, for two values of the magnetic field, $B$ = 440 and 1460 G exceeding $B_0=165$~G for $^{39}$K atom. As is seen from the right panel of Fig.\ref{fig:Fig6}, in the HPB regime four atomic transitions remain in the spectrum of $^{39}$K $D_1$ line \cite{Sargsyan3}, governed by selection rules $\Delta m_J=+1$; $\Delta m_I=0$. As the resonance linewidth in SD spectra is about 50 MHz, twice less than the spacing between transition components, they are completely spectrally resolved. The case of $B$ = 440 G corresponds to the onset of the HPB regime, and one can see that the frequency separation between the transitions labeled 3 and 4 is larger than for 1 and 2. For $B$ = 1460 G, the HPB regime is fully established, and all the components spacing becomes equidistant \cite{Sargsyan3}. We should note that the transition labeled 4 is a “guiding” transition, which is characterized by the highest (and invariable) value of probability in the group, and the probabilities of other transitions tend to this value when $B \gg B_0$ \cite{Sargsyan3}. In the case of $\sigma^-$ excitation, also four atomic transitions are observable, located far on the low-frequency range of the spectrum (not shown).

\begin{figure}[h]
\includegraphics[width=8.8cm,keepaspectratio=true]{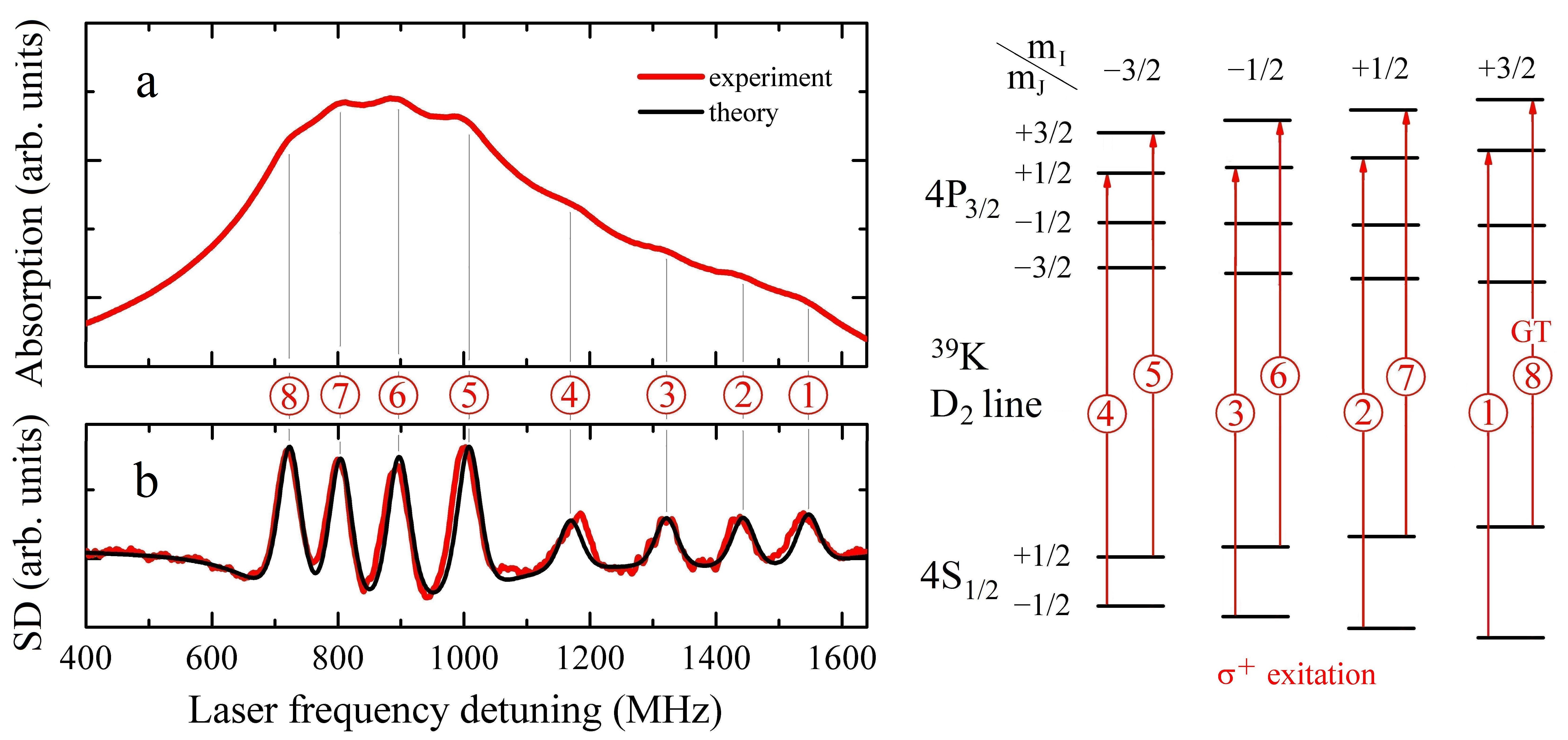}
\caption{\label{fig:Fig7} Absorption (a) and SD (b) spectra for $D_2$ line of $^{39}$K, recorded using $\sigma^+$-polarized excitation in a NC with $L\approx$ 385 nm exposed to $B$ = 500 G longitudinal magnetic field. Red lines: experiment, black line: theory. Right panel shows the transitions diagram; the eight components remaining in the HPB regime are labeled in circles.}
\end{figure}

Fig.\ref{fig:Fig7} presents absorption and SD spectra for $D_2$ line of $^{39}$K, recorded with $\sigma^+$- polarized radiation in the HPB regime ($B = 500$ G). As is seen from the right panel of the figure, eight transitions remain in the spectrum for this case (two transition groups, each containing four components). The probabilities of transitions within each group are equal, and frequency spacing of the lines is nearly the same, which is a manifestation of the HPB regime. Evidently, all the transitions are fully resolved in SD spectrum (Fig.\ref{fig:Fig7}b), unlike the raw absorption spectrum, where they are strongly overlapped. Transition labeled 8 is a “guiding” transition \cite{Klinger}. In the case of $\sigma^-$-polarized excitation, the eight atomic transitions appear far on the low-frequency range of the spectrum (not shown). 

Summarizing, we have implemented the proposed SD method for quantitative atomic spectroscopy in five particular physical problems, and have shown its effectiveness. The technique is technically simpler in realization as compared with other known methods, including saturated absorption, selective reflection using a NC \cite{Sargsyan4}. Moreover, the method preserves linear response of the media, it does not require high laser power, and does not generate extra resonances as for SA. We should mention that the SD technique can be especially advantageous for the spectroscopy of Na, where narrow resonances can be formed despite an extra-large Doppler broadening ($\sim$ 1.5~GHz). It can be also successfully implemented for studies of magnetically-induced atomic transitions and their applications demonstrated recently \cite{Tonoyan}. 

\begin{acknowledgments}
A.S.,  A.A. and  D.S. thank Armenia Science Committee for Grant No.18T-1C018. A.A. acknowledges financial support from AGBU $\&$ Pilipossian and Pilossian foundation in Geneva, as well as Foundation for Armenian Science and Technology.
\end{acknowledgments}


\end{document}